\documentstyle[12pt]{article}
\def\be{\begin{equation}}
\def\ee{\end{equation}}
\def\bea{\begin{eqnarray}}
\def\eea{\end{eqnarray}}

\author{Hans-J\"urgen Schmidt}

\title{On space--times which cannot be
distinguished by curvature 
invariants\footnote{Abstract for the 
Conference General Relativity 14,
Florence/Italy, August 6 - 12, 1995. }}

\date{}
\begin{document}
\maketitle

\centerline{Universit\"at Potsdam, Institut f\"ur Mathematik, Am
Neuen Palais 10} 
 \centerline{D-14469~Potsdam, Germany,  E-mail:
 hjschmi@rz.uni-potsdam.de}

\begin{abstract}
We present an  example that
 non--isometric 
space--times with non--vanishing curvature scalar
 cannot be distinguished by curvature invariants.
\end{abstract}

\medskip
For a positive $C^{\infty}$--function $a(u)$ let 
$$ 
ds^2 \ = \ \frac{1}{z^2} \left[ 2 \, du \, dv 
\ - \  a^2(u) \, dy^2 \ - \ dz^2 \right] \, . 
$$
In the region $z>0$, $ds^2$ represents the anti-de Sitter
space--time if and only
if $a(u)$ is linear in $u$. Now, let
$d^2a/du^2 \, < \, 0$ and
$$ 
\phi \ := \ \frac{1}{\sqrt{\kappa}} \int 
\left( - \frac{1}{a} \, \frac{d^2a}{du^2} \right) ^{1/2}
 \, du \, .
$$
Then $\Box \phi \, = \,  \phi_{,i} \, \phi^{,i} \, = \, 0$
and $R_{ij} \ - \ \frac{R}{2} \, g_{ij} \ = \ \Lambda \, 
g_{ij} \ + \ \kappa \, T_{ij}$ with 
$\Lambda \, = \, - 3$ and 
$ T_{ij} \, = \, \phi_{,i} \, \phi_{,j}$.
So $(ds^2, \, \phi)$ represents a solution of Einstein's 
equation with negative cosmological term $\Lambda $ and 
a minimally coupled massless scalar field $\phi $. 
Let $I$ be a curvature invariant of order $k$, i.e., $I$ is 
a scalar 
$$
 I = I \left( g_{ij},  \,  R_{ijlm}, \dots , R_{ijlm;i_1
 \dots \, i_k} \right)
$$
depending continuously on all its arguments. Then for
 the metric $ds^2$, $I$ does not depend on the
 function $a(u)$. 

\bigskip

This seems to be the first example that
 non--isometric 
space--times with non--vanishing curvature scalar
 cannot be distinguished by curvature invariants. The
 proof essentially uses the non--compactness of the
 Lorentz group $SO(3,1)$, 
here of the 
boosts $u \rightarrow u \, \lambda $, $v \rightarrow
  v/\lambda$. One can see this also in the
 representation theory in comparison with the
 representations of the compact rotation group $SO(4)$. 

\bigskip
 
Let $v \in R^4$ be a vector and $g \in SO(4)$ such that 
$g(v) \uparrow \uparrow v$, then $g(v)=v$. In Minkowski 
space--time $M^4$, however, there exist vectors $v \in M^4$
 and a $h \in SO(3,1)$ with $h(v)  \uparrow \uparrow v$ and
$h(v) \ne v$. This is the reason why the null frame 
components of the curvature tensor called Cartan ``scalars"
 are not always curvature scalars.

\bigskip
See H.-J. Schmidt, gr-qc/9404037, 
 Why do all the curvature invariants of a
gravitational wave vanish? in: New frontiers in gravitation,
Ed.: G. Sardanashvily, Hadronic Press, 1996, 337-344. 

\bigskip
 
I gratefully acknowledge comments by H. Baumg\"artel, 
G. Hall, A. Held, L. Herrera, K. Lake, F. de M. Paiva,
M. Rainer, and R. Schimming.

\end{document}